\documentclass[12pt]{iopart}
\pdfoutput=1

\usepackage{setstack}
\usepackage{iopams}
\usepackage{graphicx}
\usepackage{amsthm}

\begin{document}

\title{Segmentation of DNA sequences into twostate regions and melting fork regions}
\author{Eivind T{\o}stesen$^{1,2}$, Geir Kjetil Sandve$^3$, Fang Liu$^{4,5}$ and Eivind Hovig$^{1,3,4}$}
\address{$^1$ Department of Medical Informatics, Norwegian Radium Hospital, N-0310 Oslo, Norway}
\address{$^2$ Department of Mathematics, University of Oslo, N-0316 Oslo, Norway}
\address{$^3$ Department of Informatics, University of Oslo, N-0316 Oslo, Norway}
\address{$^4$ Department of Tumour Biology, Institute for Cancer Research, Norwegian Radium Hospital, N-0310 Oslo, Norway}
\address{$^5$ PubGene AS, Vinderen, N-0319 Oslo, Norway}
\ead{eivindto@math.uio.no}

\begin{abstract}
The accurate prediction and characterization of DNA melting domains by
computational tools could facilitate a broad range of biological
applications. However, no algorithm for melting domain prediction
has been available until now.
The main challenges include the difficulty of mathematically
mapping a qualitative description of DNA melting domains
to quantitative statistical mechanics models,
as well as the absence of 'gold standards'
and a need for generality.
In this paper,
we introduce a new approach to identify the twostate regions and
melting fork regions along a given DNA sequence. Compared with an \emph{ad
hoc} segmentation used in one of our previous studies, the new algorithm
is based on boundary probability profiles, rather than standard melting
maps. We demonstrate that a more detailed characterization of the DNA
melting domain map can be obtained using our new method, and this 
approach is independent of the choice of DNA melting model.
We expect this work to drive our understanding of DNA melting domains one step further.
\end{abstract}
\submitto{\JPCM}


\maketitle

\section{Introduction}
The organisation of \emph{DNA melting domains} is
the main characteristic of DNA melting cooperativity.
However, a mathematical definition of melting domains and an algorithm for locating
them in a given sequence does not exist. 

It has been known since the 1970s that DNA melts stepwise, domain after domain,
in a reproducible series of subtransitions as temperature is increased
\cite{WartellBenight}.
DNA melting domains have also been called cooperatively melting regions (CMR)
\cite{gotoh83prediction},
thermalites
\cite{WartellBenight},
cooperative units, and isomelting domains
\cite{ekstrom2007}.
DNA melting domains are regions of tens to hundreds of basepairs,
with locations determined by the sequence.
The melting of each domain is approximately a \emph{twostate transition}.
In the strict sense,
twostate means that only two states (``11111'' and ``00000'') are involved,
no intermediate states are populated,
all basepairs thus open and close in unison,
and the melting domains are the smallest units of melting.
Short DNAs and oligonucleotides usually consist of only one domain.
The stability and melting temperature of a domain not only
depends on its internal sequence composition,
but also on the stabilities of adjacent domains.

A distinction must be made between domains and bubbles:
The melting domains indicate potential melting events, such as the creation,
growth or merging of bubbles. Melting domains are regions with fixed boundaries,
while helix-coil boundaries fluctuate and depend on temperature.
Various approaches to computing bubbles already exist
\cite{tostesen_amb2008,TobiasAmbjornsson04152007,erp:218104}.

DNA melting can be predicted by various statistical mechanics
models and algorithms
\cite{PS,DPB93,15579230}.
The formulation of these models involves microscopic interactions
and variables at the basepair level,
but the concept of melting domains is absent in these formulations.
There is no explicit representation of melting domains in
the input or the output of these algoritms.
Neither are melting domains simply due to certain input parameters,
such as a bubble nucleation barrier,
but are rather emergent properties that depend on all the parameters
and the whole sequence.
The concept of melting domains comes into play at a postprocessing stage
and in the interpretation of the output.
For example, plots of melting curves and melting maps
usually contain some characteristic features, such as peaks, steps or plateaux,
that may be associated with melting domains.
In some studies, this association is only qualitative,
because a precise definition of melting domains was not made.
However, in the analysis and decomposition of optical melting curves,
a quantitative relationship between peaks and domains is assumed
\cite{yen80analysis,blake98thermal}.

The concept of DNA melting domains rests on
work by Azbel in the late 1970s
\cite{azbel80dna-sequencing1,azbel80dna-sequencing2}.
He described a method for determining the groundstate of DNA
as a function of temperature.
His analysis showed that a sequence is partitioned into \emph{groundstate domains}.
Each groundstate domain has a temperature at which
the groundstate changes from helix to coil.
These temperatures define the order in which the groundstate changes
with increasing temperature.
He proposed the groundstate versus temperature scenario
as a simplified model of the DNA melting process
to provide quantitative predictions.
The predictive power of his groundstate model soon turned out to be limited,
with the conclusion that all microstates must be included
in the partition function
\cite{lyubchenko-yu82a-comparison,gotoh83prediction}.
In spite of that, the groundstate model seems to have strongly influenced,
if not dictated, the concept of DNA melting domains. 
In this study, we consider
melting domains and groundstate domains to be different concepts.

\begin{figure}
 \includegraphics[width=13.5cm]{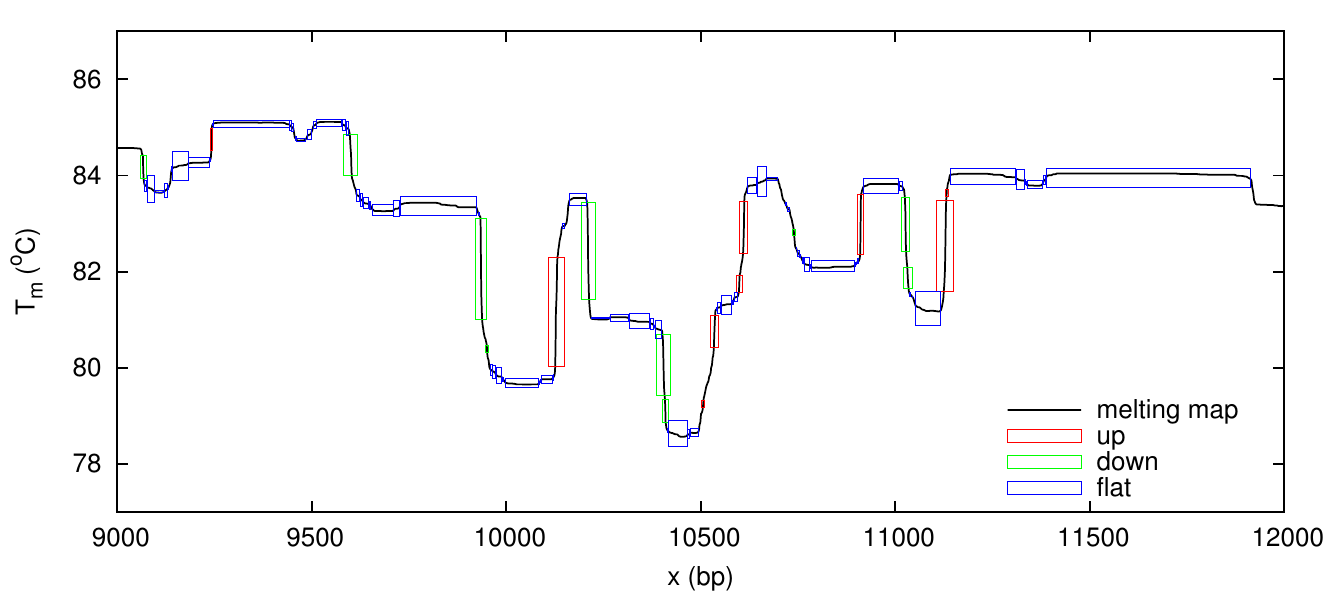}%
 \caption{
The \emph{ad hoc} segmentation method that was
applied to the human genomic melting map
\cite{liu2007}.
The boxes indicate the up, down and flat regions that
are determined by the local slope of the melting map.
\label{fig1}
}
\end{figure}

In a previous study, we approached the problem of computing melting domains.
In a bioinformatical analysis of the human genomic melting map,
we found it useful to extract qualitative features of the curve,
such as flat plateaux and steep slopes
\cite{liu2007}.
We devised an \emph{ad hoc} method
to do such a segmentation of the melting map,
illustrated in \fref{fig1},
with threshold values that were chosen intuitively.
Using this method, we identified melting domains as \emph{flat segments},
based on the assumption that consecutive basepairs melt in unison
if they have equal melting temperatures.
However, this is not necessarily true.
We only know that the reverse implication is true:
basepairs share the same melting temperature if they melt together.
In this work, we seek a more proper way of doing it.

We consider ``twostateness'' to be the defining property of melting domains.
Previous tradition has claimed that twostateness can be detected at the
calorimetric level and by considering the widths of transitions.
But counter examples have shown that those approaches were error-prone
\cite{zhou99the-calorimetric,bakk04is-it-always}.
Instead, we consider the preferred locations of melting forks.
We describe a segmentation method based on \emph{boundary probability} profiles.
Already in his 1974 paper, Poland discussed the 
probabilities that a 01 or 10 boundary exists at a specific unit
\cite{poland}.
This seems to have had little impact, however.
Most studies have used the standard \emph{probability profiles}---even when
melting forks was the main interest
\cite{carlon:178101,carlon:051916}.
The present method produces segmentations in which melting domains do not
cover the whole sequence, the result is not simply a partitioning.
Instead, we find that there are \emph{twostate regions}
interspersed with \emph{melting fork regions}.

Many cellular functions require local
single-strandedness of DNA, for example,
DNA replication and gene regulation. A number
of DNA-associated proteins contribute
to the information capacity of the
four-base DNA code
\cite{vonhippel,chaires}.
Thus, for instance nuclear
matrix proteins and nucleosomes  are part of the
chromatin level of information, and participate
to modulate trancription  factor activity.
Ultimately, the DNA sequence itself is
insufficient to understand the dynamic interplay
of the molecules, as indicated by the performance
of prediction algorithms for the binding of
these types of molecules along the genomic DNA.
These algorithms are presently mostly based on
the sequence level of information. In order to
advance further in the understanding of these
aspects, it appears necessary to incorporate the
physical properties of DNA in a comprehensive
way, and we here attempt to precisely define the
extent of the genomic features of DNA melting.
\section{Methods}
There are a number of alternative DNA melting models
being used by various research groups.
The present work aims at generality: The concepts and methods
should be applicable to each of the DNA melting models.
Our approach is formulated solely in terms of equilibrium probabilities
that should be computable with any model, without reference to quantities or properties
pertaining to one model only.
In some DNA melting models, the state of a basepair is a binary variable (0 or 1),
where 0 is the melted/coil/open state and 1 is the bound/helix/closed state,
while in other DNA melting models, a state can be a distance and/or angle.
In the latter models, however, a criterion for distinguishing single-stranded (ss)
and double-stranded (ds) is sometimes applied.

\begin{figure}
 \includegraphics[width=8.6cm]{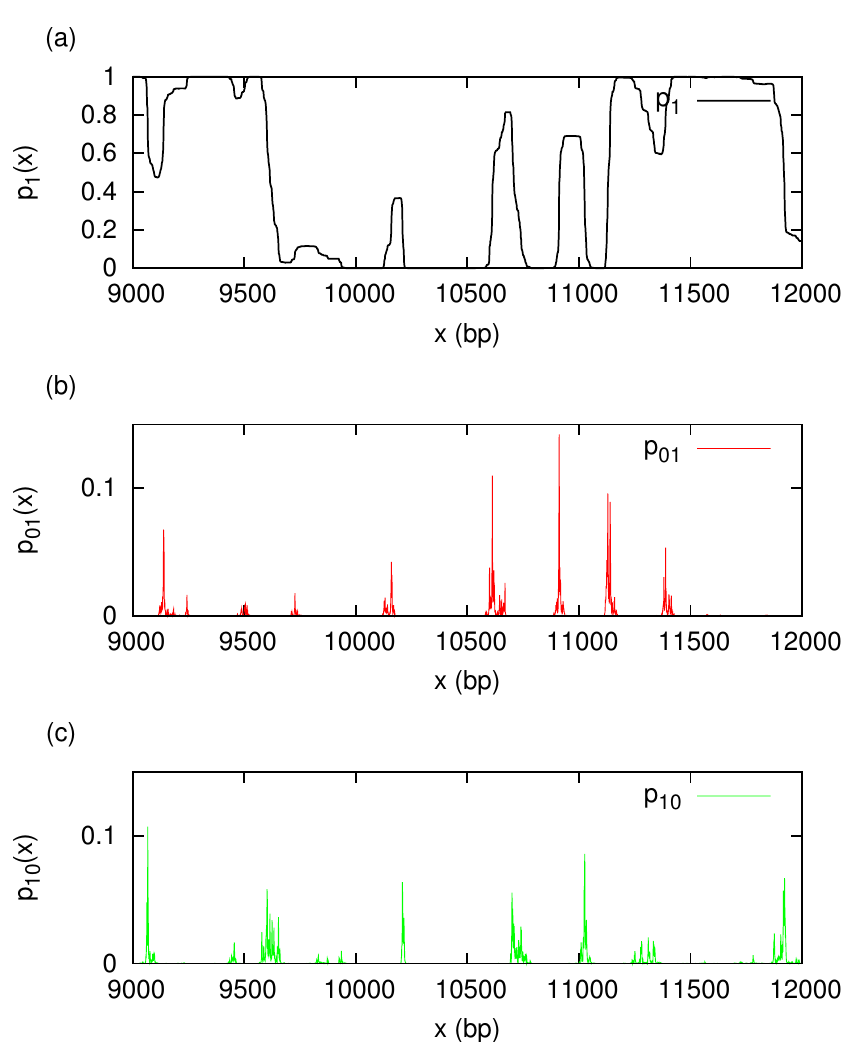}%
 \caption{
The (a) probability profile (in black), (b) 01 boundary probability profile (in red), and (c) 10 boundary
probability profile (in green), of the human mitochondrion sequence
calculated at $T=83.678^{\circ }\text{C}$ and $[\text{Na}^{+}]=0.075\text{ M}$
are viewed in the window 9000 bp--12000 bp as an example.
(Colour mnemonic: green=bubble start, red=bubble stop.)
\label{fig2}
}
\end{figure}

We assume that the bases in the sequence are numbered $1,\ldots,N$
and that each conformation (microstate) may be represented as a chain of 0 and 1 values.
At each base position $x$, we define the probability of a 0 or 1:
\numparts
\begin{eqnarray}
p_{0}(x)&=&P(\text{\ldots X}\stackrel{x}{0}\text{X\ldots})\\
p_{1}(x)&=&P(\text{\ldots X}\stackrel{x}{1}\text{X\ldots}).
\end{eqnarray}
\endnumparts
In these equations,
1 is a bound basepair (ds),
0 is a melted basepair (ss),
X indicates either 0 or 1,
and the sequence position $x$ is indicated.
At each specific temperature, we can calculate a \emph{probability profile},
which is the $p_{1}(x)$-values for $1\leq x\leq N$.
A \emph{melting map} can be interpolated from several probability profiles and indicates at each position
the temperature at which $p_{1}(x)=\frac12$.
Probability profiles and melting maps have been standard tools in biological applications.
The present approach employs another set of probabilities.
Each segment of two nearest neighbour base positions $[x-1,x]$
has four possible conformations. Accordingly, we consider the probabilities
\numparts
\begin{eqnarray}
p_{00}(x)=P(\text{\ldots X0}\stackrel{x}{0}\text{X\ldots})\\ 
p_{01}(x)=P(\text{\ldots X0}\stackrel{x}{1}\text{X\ldots})\\
p_{10}(x)=P(\text{\ldots X1}\stackrel{x}{0}\text{X\ldots})\\ 
p_{11}(x)=P(\text{\ldots X1}\stackrel{x}{1}\text{X\ldots})
\end{eqnarray}
\endnumparts
for $2\leq x\leq N$.
$p_{01}(x)$ and $p_{10}(x)$ are called \emph{boundary probabilities}.
We assume that they can be calculated to obtain \emph{boundary probability profiles}
at each temperature. 
\Fref{fig2}
shows an example of 01 and 10 boundary probability profiles compared with the
probability profile.
By basic probability laws, we have:
\begin{eqnarray}
p_{0}(x)+p_{1}(x)=1\\
p_{00}(x)+p_{01}(x)+p_{10}(x)+p_{11}(x)=1
\end{eqnarray}
and
\numparts
\begin{eqnarray}
p_{00}(x)=p_{0}(x-1)p_{0|0}(x)\\
p_{01}(x)=p_{0}(x-1)p_{1|0}(x)\\
p_{10}(x)=p_{1}(x-1)p_{0|1}(x) \label{condprobs}\\
p_{11}(x)=p_{1}(x-1)p_{1|1}(x)
\end{eqnarray}
\endnumparts
Here we have introduced the \emph{conditional probabilities},
for example, the probability $p_{1|0}(x)$ that there is a 1 at position $x$
given that there is a 0 at position $x-1$.
\subsection{Twostate regions}
In this section, we propose a formal definition of twostateness.
Usually, the adjective \emph{twostate} indicates a property of transitions,
but we will use it to indicate a property of sequence regions.
Literally, \emph{twostate} means that only two possible conformations of a region $[x,y]$,
the all 1's and all 0's, can exist at any instant and at any temperature.
This ``all-or-none'' situation may be expressed as
\begin{equation}
\label{strictly}
P(\text{\ldots X}
\underbrace{\stackrel{x}{1}\cdots \stackrel{y}{1}}_{\text{all 1's}}
\text{X\ldots})+
P(\text{\ldots X}
\underbrace{\stackrel{x}{0}\cdots \stackrel{y}{0}}_{\text{all 0's}}
\text{X\ldots})
=1,
\end{equation}
or, equivalently, that there is zero probability at all temperatures
of the region containing any ds/ss boundaries.
However, regions that are strictly twostate do not exist.
We expect that there is always a nonzero probability of
partially melted intermediates.
A more realistic definition must incorporate
a small probability of thermally induced ``defects''
in a twostate region.
We achieve this by means of a threshold $\epsilon>0$:
A region $[x,y]$ is called \emph{globally twostate} if
\begin{equation}
\label{globally}
P(\text{\ldots X}
\underbrace{\stackrel{x}{1}\cdots \stackrel{y}{1}}_{\text{all 1's}}
\text{X\ldots})+
P(\text{\ldots X}
\underbrace{\stackrel{x}{0}\cdots \stackrel{y}{0}}_{\text{all 0's}}
\text{X\ldots})
\geq 1-\epsilon
\end{equation}
at all temperatures. The term \emph{globally}
refers to the above two probabilities, in which
the conformation of the whole region $[x,y]$ is specified.
Such probabilities may be computationally challenging. 
We expect that computing globally twostate regions
could involve high algorithmic complexities.
For this practical reason, we will instead study regions that are \emph{locally} twostate.
\Eref{globally} applied locally to a nearest neighbour segment at position $x$
can be written
\begin{equation}
p_{11}(x)+p_{00}(x)\geq 1-\epsilon
\end{equation}
or, equivalently,
\begin{equation}
p_{01}(x)+p_{10}(x)\leq \epsilon.
\end{equation}
However, in the following definition we
will require each of $p_{01}(x)$ and $p_{10}(x)$, rather than their sum,
to be smaller than $\epsilon$.
A region $[x_{\text{start}},x_{\text{end}}]$ is \emph{locally twostate} if
for each $x\in [x_{\text{start}}+1,x_{\text{end}}]$:
\numparts
\begin{eqnarray}
\label{locally}
p_{01}(x)\leq \epsilon\\
p_{10}(x)\leq \epsilon
\end{eqnarray}
\endnumparts
at all temperatures.
A region being globally twostate implies that it is locally twostate,
but not vice versa. However, if region $[x,y]$ is locally twostate, then
(by Boole's inequality):
\begin{equation}
P(\text{\ldots X}
\underbrace{\stackrel{x}{1}\cdots \stackrel{y}{1}}_{\text{all 1's}}
\text{X\ldots})+
P(\text{\ldots X}
\underbrace{\stackrel{x}{0}\cdots \stackrel{y}{0}}_{\text{all 0's}}
\text{X\ldots})
\geq 1-2n\epsilon
\end{equation}
at all temperatures, where $n=x_{\text{end}}-x_{\text{start}}$,
that is, the region is globally twostate with a less strict threshold value.
\subsection{Boundary probability segmentation}
So far, we have proposed global and local variants of twostate regions,
but we have not indicated how to determine
the start and end positions of such regions.
For globally twostate regions, we leave this as an open question.
For locally twostate regions, the starts and ends follow naturally when
evaluating at each position from 2 to $N$ whether or not
each of the two boundary probabilities is small enough
at all temperatures. This yields four possible cases at each position $x$:
\begin{enumerate}
\item $\forall T: p_{01}(x)\leq \epsilon\text{ and }\forall T: p_{10}(x)\leq \epsilon$ (twostate)
\item $\exists T: p_{01}(x)> \epsilon\text{ and }\forall T: p_{10}(x)\leq \epsilon$ (01 fork)
\item $\forall T: p_{01}(x)\leq \epsilon\text{ and }\exists T: p_{10}(x)> \epsilon$ (10 fork)
\item $\exists T: p_{01}(x)> \epsilon\text{ and }\exists T: p_{10}(x)> \epsilon$ (double fork)
\end{enumerate}
By thus classifying each nearest neighbour segment $[x-1,x]$, we obtain a segmentation
that divides the sequence into four types of regions.
For each region $[x_{\text{start}},x_{\text{end}}]$, one of the cases (i)--(iv) is true
at all internal positions $x\in [x_{\text{start}}+1,x_{\text{end}}]$,
but not true at the flanking positions $x_{\text{start}}$ and $x_{\text{end}}+1$.
A case (i) region is locally twostate, hereafter referred to as a \emph{twostate region}.
In a case (ii) region, there are large probabilities of 01 boundaries at some temperatures,
while the 10 boundary probabilities always remain small. This is called a \emph{01 fork region}.
Likewise, in a case (iii) region, there are large probabilities of 10 boundaries at some temperatures,
while the 01 boundary probabilities remain small always. This is called a \emph{10 fork region}.
The latter two types of regions are called \emph{melting fork regions}.
The last case, (iv), is a region in which there may be large probabilities
of both 01 and 10 boundaries. This is called a \emph{double fork region}.

\begin{figure}
 \includegraphics[width=13.5cm]{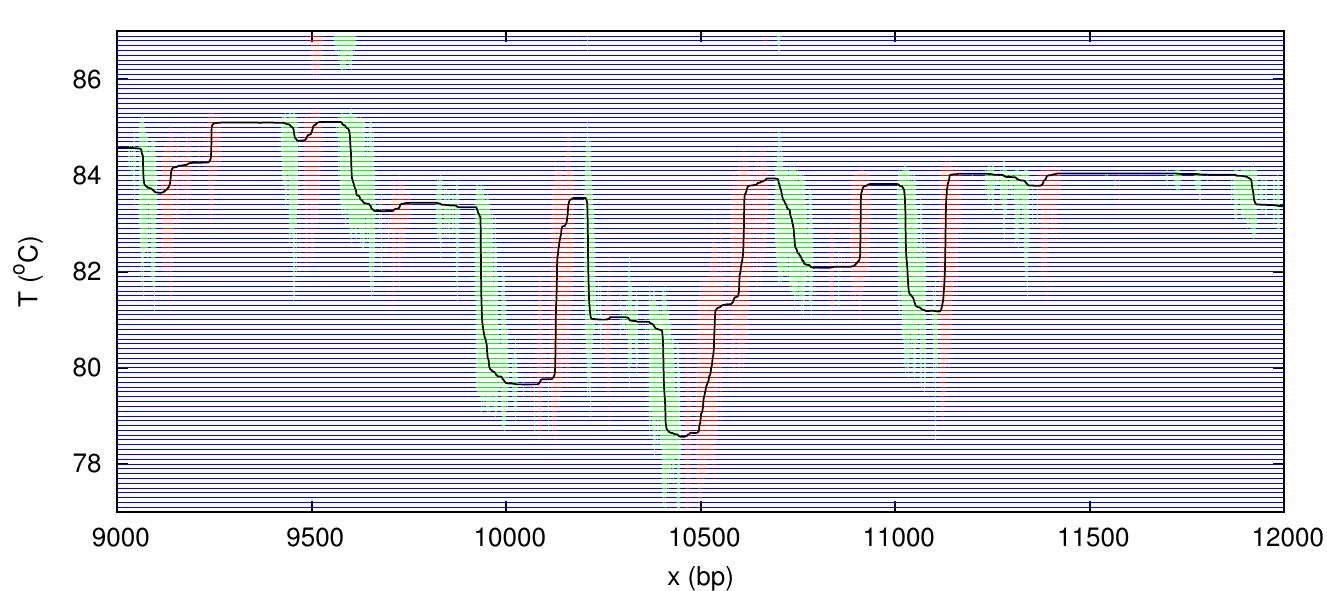}%
 \caption{
At each temperature (in steps of $0.1^{\circ }\text{C}$) and
at each position $x$ in the sequence, the colors red, green, blue
indicate the presence of a
01 boundary ($p_{01}(x)> \epsilon$ and $p_{10}(x)\leq \epsilon$),
10 boundary ($p_{01}(x)\leq \epsilon$ and $p_{10}(x)> \epsilon$),
or no boundary ($p_{01}(x)\leq \epsilon$ and $p_{10}(x)\leq \epsilon$)
in the parameterfree case: $\epsilon=\epsilon_{\text{min}}$.
Boundaries tend to ``live'' during temperature intervals up to $4^{\circ }\text{C}$ wide.
On top of the colour map is plotted the melting map (black curve).
Note that the two were calculated
independently from each other.
\label{fig3}
}
\end{figure}

\Fref{fig3}
illustrates that each point $(x,T)$ in the plane can be classified and given a colour according to the comparison of the
two boundary probabilities with $\epsilon$. In this plot,
twostate regions correspond to entirely blue vertical columns,
while melting fork regions are columns that contain
red or green ``islands''.
\Fref{fig3} also shows how some of the islands correspond to
the steep slopes on the melting map.
\subsection{Sampling algorithm}
An algorithm for finding twostate regions and melting fork regions
must evaluate statements of the type ``for all temperatures''.
An approximate method is to sample only a finite, but representative set of temperatures,
for example, by scanning a range from $T_{\text{low}}$ to $T_{\text{high}}$
with incremental steps $\Delta T$.
We use values $T_{\text{low}}$ and $T_{\text{high}}$ corresponding
to helicities $\theta =0.999$ and $\theta =0.001$, so as to
cover most of the subtransitions.
This usually means $T_{\text{high}}-T_{\text{low}}\approx 20^{\circ }\text{C}$.
The choice of step size is a trade-off.
The smaller the $\Delta T$, the more temperatures are taken into account
and the more accurate is the segmentation, but at the cost of longer computation time.
Melting forks only exist over a temperature interval (\fref{fig3}),
so the resolution must be high enough to detect the
shortest of such temperature intervals.

\begin{figure}
 \includegraphics[width=8.5cm]{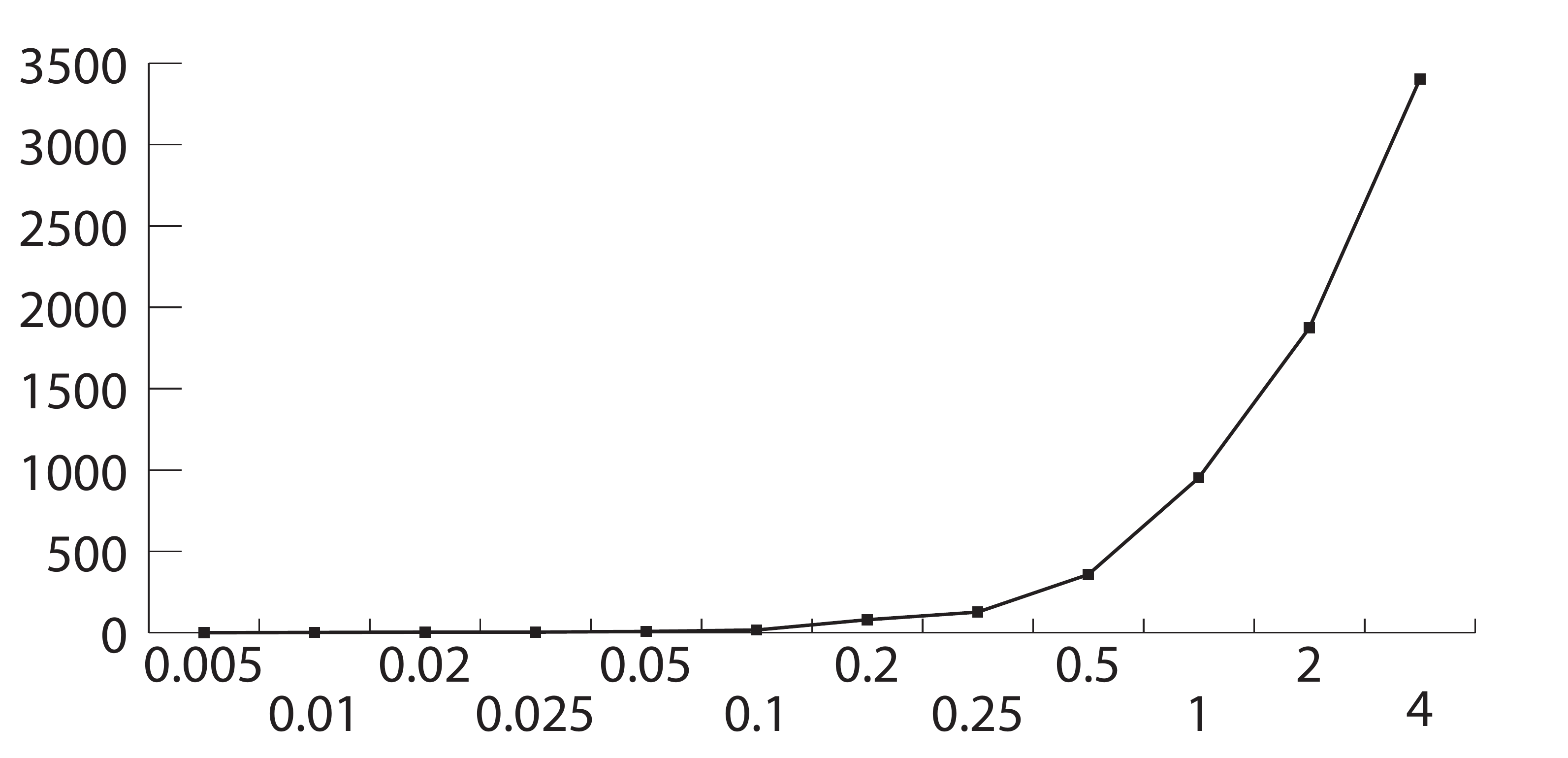}%
 \caption{
Convergence test for the sampling algorithm, using a segmentation based on step size of $0.005^{\circ }\text{C}$ as reference. Different choices of step size are given on the x-axis, with the y-axis denoting Hamming distance between resulting segmentations and the reference segmentation.
\label{fig4}
}
\end{figure}

\Fref{fig4}
shows how the segmentation depends on sampling resolution. 
A segmentation based on a very small step size of $0.005^{\circ }\text{C}$ is used as reference, and the figure shows the Hamming distance between this reference and segmentations based on larger step size. The hamming distance measures the number of positions that differs with respect to the class (twostate and melting forks) between two compared segmentations.

As can be seen from the figure, the use of step sizes above $1^{\circ }\text{C}$ alters the resulting segmentation substantially, while the segmentation converges for step sizes below $0.1^{\circ }\text{C}$. Based on this analysis we have chosen $\Delta T=0.1^{\circ }\text{C}$ as standard, since this gives a segmentation almost identical to segmentations achieved with smaller step sizes and longer computational time.

\subsection{Determining the $\epsilon$-value}
The parameter $\epsilon$ is the threshold used for distinguishing
large from small boundary probabilities. There are two ways of determining its
value:
(1) as an arbitrary input to the algorithm and
(2) as a parameter-free output of the algorithm.

\begin{figure}
 \includegraphics[width=11cm]{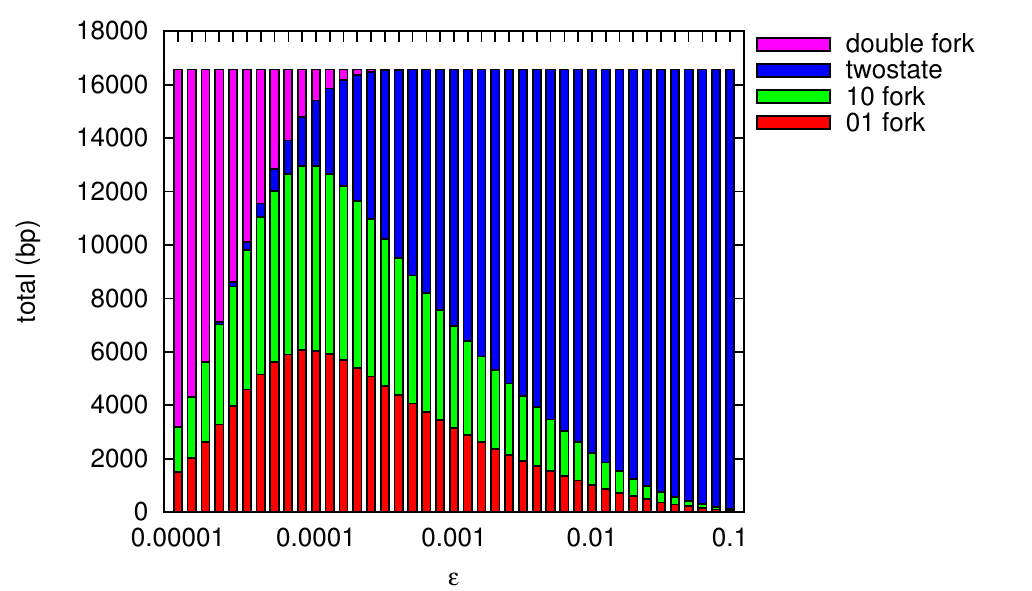}%
 \caption{
Stacked histogram showing the total amounts of twostate regions (blue),
01 fork regions (red), 10 fork regions (green), and double fork regions (magenta)
in a 16571 bp sequence, plotted over a range of $\epsilon$-values (logarithmic axis).
The double fork regions disappear at $\epsilon_{\text{min}}\approx 0.0006$.
\label{fig5}
}
\end{figure}

Thermal fluctuations create boundaries anywhere in the sequence,
which imposes a background noise level
in the boundary probability profiles. On the other hand,
the sequence encodes the preference for certain boundary locations,
which gives rise to high peaks above the background noise (\fref{fig2}).
The parameter-free $\epsilon$-value separates the peaks from the noise.
\Fref{fig5}
shows how the segmentation depends on $\epsilon$. At $\epsilon=0.1$, the boundary
probabilities are considered small almost everwhere, there are no melting fork regions.
At the other extreme, $\epsilon<0.00001$, the boundary
probabilities are considered large almost everwhere, there are no twostate regions
and most of the sequence is considered to be double fork. We interpret
the double fork regions as a reflection of the background noise.
We define the parameterfree $\epsilon_{\text{min}}$
as the value above which the double fork regions disappear.
The $\epsilon_{\text{min}}$ depends on the sequence and can be written
\begin{equation}
\label{epsilonmin}
\epsilon_{\text{min}}=\max_{T}[\max_{x}(\min\{p_{01}(x),p_{10}(x)\})].
\end{equation}
To avoid detecting thermal noise, one should use an $\epsilon\geq\epsilon_{\text{min}}$.
In parameter-free segmentations, we do a sampling over a temperature interval
$[T_{\text{low}},T_{\text{high}}]$ to determine the $\epsilon_{\text{min}}$
by \eref{epsilonmin}.
\Fref{fig5} shows that the amounts of twostate, 01 fork, and 10 fork regions
are about a third each at $\epsilon_{\text{min}}$. 
\subsection{Approximations derived from probability profiles and melting maps}
\Fref{fig2} shows a correspondence between a probability profile and
the two boundary probability profiles: Steep increases in $p_1(x)$
coincide with peaks in $p_{01}(x)$, while steep decreases
coincide with peaks in $p_{10}(x)$. This suggests that
the twostate, 01 fork, and 10 fork regions can be estimated approximatively
from standard probability profiles alone.

For $2\leq x\leq N$ we can define the \emph{slope} of a probability profile
as the difference:
\begin{equation}
\Delta p_1(x)=p_1(x)-p_1(x-1).
\end{equation}
By
\numparts
\begin{eqnarray}
p_{1}(x)=p_{01}(x)+p_{11}(x)\\
p_{1}(x-1)=p_{10}(x)+p_{11}(x)
\end{eqnarray}
\endnumparts
we derive
\begin{equation}
\label{p01-p10}
\Delta p_1(x)=p_{01}(x)-p_{10}(x).
\end{equation}
This equation explains the correspondence in \fref{fig2}
between the slope in the probability profile and the peaks in the
boundary probabilities.
It also shows that there is a loss of information when two boundary probability
profiles are ``collapsed'' into the probability profile.
Instead of classifying each nearest neighbour segment $[x-1,x]$
according to the cases (i)--(iv) above, we could do the following classification:
\begin{equation}
\label{ppseg}
\text{class}(x)=\cases{
\text{up}&if $\exists T: \Delta p_1(x)>\epsilon$\\
\text{flat}&if $\forall T: |\Delta p_1(x)|\leq \epsilon$\\
\text{down}&if $\exists T: \Delta p_1(x)<-\epsilon$\\
}
\end{equation}
(Note that this segmentation of the probability profiles into
up, down, and flat regions is not the same as the
segmentation of the melting map.)
From \eref{ppseg} and \eref{p01-p10} it follows that
each twostate region is contained inside a flat region.
Furthermore, if $\epsilon\geq\epsilon_{\text{min}}$
(no double forks), then
each up region is contained inside a 01 fork region and
each down region is contained inside a 10 fork region.
Used as an approximation, \eref{ppseg} would
misclassify some (parts of) melting fork regions
as being twostate. 

We have not analytically solved how these results translate
to the melting map segmentation \cite{liu2007}, which has the form:
\begin{equation}
\label{mmseg}
\text{class}(x)=\cases{
\text{up}&if $\Delta T_{\text{m}}(x)>\epsilon_1$\\
\text{flat}&if $|\Delta T_{\text{m}}(x)|\leq\epsilon_2$\\
\text{down}&if $\Delta T_{\text{m}}(x)<-\epsilon_1$\\
\text{none}&otherwise\\
}
\end{equation}
where $\Delta T_{\text{m}}(x)=T_{\text{m}}(x)-T_{\text{m}}(x-1)$.
Note that this has two thresholds: $\epsilon_1$ and $\epsilon_2$.
The melting map segmentation with $\epsilon_1=0.13$ and $\epsilon_2=0.01$
is illustrated in \fref{fig1}.
\subsection{Subtransitions and fork-fork types}
Once a segmentation is produced, it is possible to distinguish
different types of twostate regions, called \emph{fork-fork} types,
by identifying the nearest
flanking or bracketing melting fork region on each side.
If a twostate region is close to the sequence end (5' or 3'), then
it is possible that no melting fork regions exist before that end.
On each side, therefore, there are three possibilities (01, 10, either 5' or 3'),
which combines to
nine possible fork-fork types of twostate region.
We write the fork-fork types in the notation
01\_10, 10\_10, 5'\_10, 01\_3', etc.
The nine fork-fork types are listed in \tref{tab}.

In the human genomic melting map, we divided the flat segments into
four types called \emph{top, bottom, upstair, and downstair}
\cite{liu2007}: We compared the melting map temperature averaged over
a flat segment with the melting map temperatures averaged over
the two flanking regions of equal length. The flat segment is a top if both flanking
temperatures are lower, it is a bottom if both are higher, it is
an upstair if leftside is lower and righthand side is higher, and downstair otherwise.
\Tref{tab}
\Table{\label{tab} Twostate regions are divided into nine \emph{fork-fork} types.
Four of these correspond to \emph{meltmap} types of flat regions:
top, bottom, upstair, and downstair
\cite{liu2007}.
The \emph{subtransition} indicates the most likely melting process of each region
\cite{yen80analysis,meltsim},
based on Azbel's distinction according to the
change $n_{\text{b}}$ in the number of boundaries
\cite{azbel80dna-sequencing1}.
The \emph{effect on} $T_{\text{m}}$
is the deviation from the expectation based on GC\% content
\cite{azbel80dna-sequencing1}.}
\br
Fork-fork & Meltmap & Subtransition & $n_{\text{b}}$ & Effect on $T_{\text{m}}$\\
\mr
01\_10 & top & V & -2 & $T_{\text{m}}<T_{\text{GC\%}}$\\
10\_01 & bottom & I & +2 & $T_{\text{m}}>T_{\text{GC\%}}$\\
01\_01 & upstair & III & 0 & $T_{\text{m}}\approx T_{\text{GC\%}}$\\
10\_10 & downstair & III & 0 & $T_{\text{m}}\approx T_{\text{GC\%}}$\\
5'\_10 & --- & IV & -1 & $T_{\text{m}}<T_{\text{GC\%}}$\\
5'\_01 & --- & II & +1 & $T_{\text{m}}\approx T_{\text{GC\%}}$\\
01\_3' & --- & IV & -1 & $T_{\text{m}}<T_{\text{GC\%}}$\\
10\_3' & --- & II & +1 & $T_{\text{m}}\approx T_{\text{GC\%}}$\\
5'\_3' & --- & --- & $0$ & $T_{\text{m}}\approx T_{\text{GC\%}}$\\
\br
\endTable
shows the correspondence between the four melting map types
and four of the fork-fork types. Because of the lengths of the chromosomes,
we did not consider the other five types of flat segment at the sequence ends.
We found that the type of a flat segment (top and bottom) has a strong effect on
its melting temperature, that is, the stability depends on neighbouring domains,
not only the internal GC content
\cite{liu2007}.

These findings were in accordance with Azbel's prediction of the deviation of
a domain's $T_{\text{m}}$ from the GC\%-based Marmur-Doty prediction
\cite{azbel80dna-sequencing1,yen80analysis,gotoh83prediction,WartellBenight}.
He distinguished five types of subtransition of a melting domain by the number of boundaries
they create: (I) nucleation of a bubble, (II) unzipping from an end, (III) internal growth of
a bubble, (IV) merging of a bubble and an unzipping end, and (V) merging of two bubbles.

\Tref{tab} shows that the fork-fork classification corresponds
to distinguishing the left and right variants (mirror images) of
the subtransition types II, III and IV.
The 5'\_3' type corresponds to oligonucleotides that do not contain any melting fork regions
and melt and dissociate in a single step. The table summarizes the expected relationships between the different types of regions and
transitions. For example, when a 01\_3' region melts, it will most likely open to
merge with an existing bubble to the left and with the 3' end to the right. Otherwise,
it would create a 10 fork to the left and/or a 01 fork to the right, in contradiction with
its type being 01\_3'. The 01\_3' region is destabilized by the flanking bubble on the left,
lowering its $T_{\text{m}}$.
\section{Results}
The methodology for determining melting domains is in itself the main result of this study. Although melting domains have been used as a concept for a long time, this paper presents a principled approach to calculation of the domains. 
With the melting segmentation produced by the methodology as a proposed reference, we here analyze melting segmentation of biological DNA sequences and compare our proposed reference to the previous \emph{ad hoc} method for the prediction of melting domains.

We first give a direct visualization of melting domains in a selected portion of the mitochondrion, and then provide some statistical results from the full human chromosome 21 segmentation.

\begin{figure}[htbp]
   \centering
   \includegraphics[width=6in]{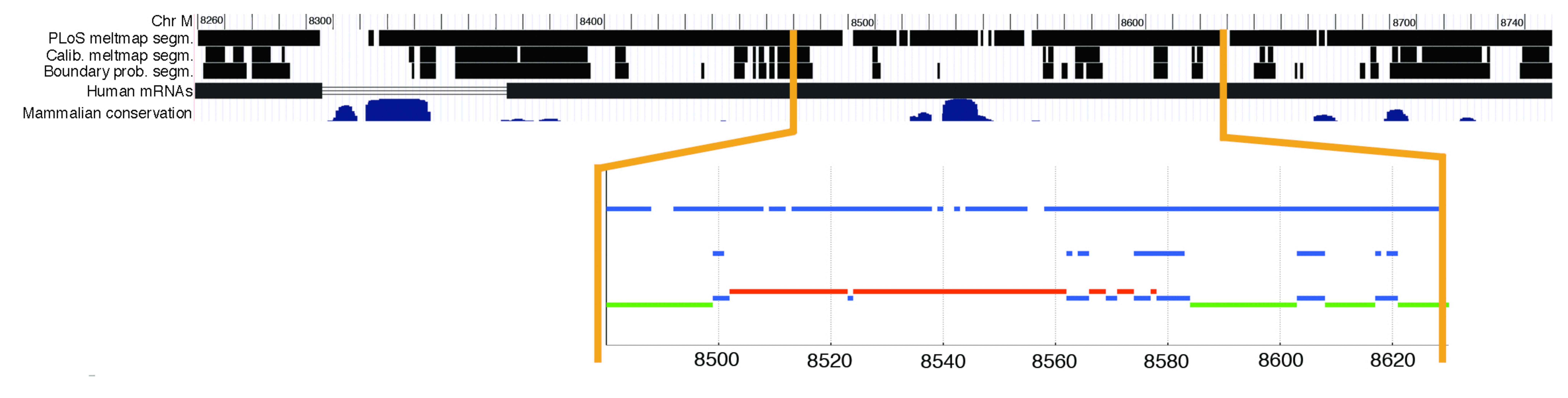} 
   \caption{Visualization of twostate and melting fork regions in the UCSC genome browser. Position 8250 to 8750 of the mitochondrial genome is shown.
   Also shown is a genomic type track, in this case
evolutionary conservation of the genomic
sequence. Although there is no direct information
to be obtained from this small segment of DNA,
these two features, or other genomic features,
may be compared to the information provided by
the segmentations, in order to uncover hidden
dependencies.
   A zoomed in version is also shown for a part of the segmentation, with twostate regions denoted in blue, and ``01'' and ``10'' forks further specified in red and green respectively.
 }
   \label{fig6}
\end{figure}

\Fref{fig6}
shows the segmentation from position 8250 to 8750 of the mitochondrial genome. Black bars denote twostate regions, while areas between bars denote melting fork regions.
As can be seen from the figure, the segmentation typically forms several tight clusters of twostate regions. Very short melting fork regions are typically occurring between twostate regions within each such cluster, while one or a few long melting fork regions are separating the clusters. Also, there are a few very short, isolated twostate regions.

The segmentation is further compared to the melting map segmentation
\cite{liu2007}. Flat regions of this segmentation should correspond to twostate regions of the new segmentation and are denoted in black. Non-flat regions correspond to melting fork regions, with ``up'' regions corresponding to ``01'' forks and ``down'' regions corresponding to ``10'' forks . As can be seen from the figure, the \emph{ad hoc} procedure used in the previous paper gave rise to a much larger amount of flat regions. The \emph{ad hoc} segmentation typically gave very long flat regions, only separated by short non-flat regions. 

As twostate (flat) regions covered a much larger part of the genome according to the previous segmentation procedure, we also constructed a third segmentation. This segmentation was based on the melting map segmentation \eref{mmseg}, but with parameter values $\epsilon_1=0.13$ and $\epsilon_2=0.0002$ calibrated to give similar global amounts of twostate regions
and flat segments. This calibrated melting map segmentation resembles the new segmentation, although some minor differences are noticeable. For instance, where one segmentation assigns a continuous twostate region, the other segmentation sometimes assigns multiple clustered regions. 

\begin{figure}[htbp]
   \centering
   \includegraphics[width=5in]{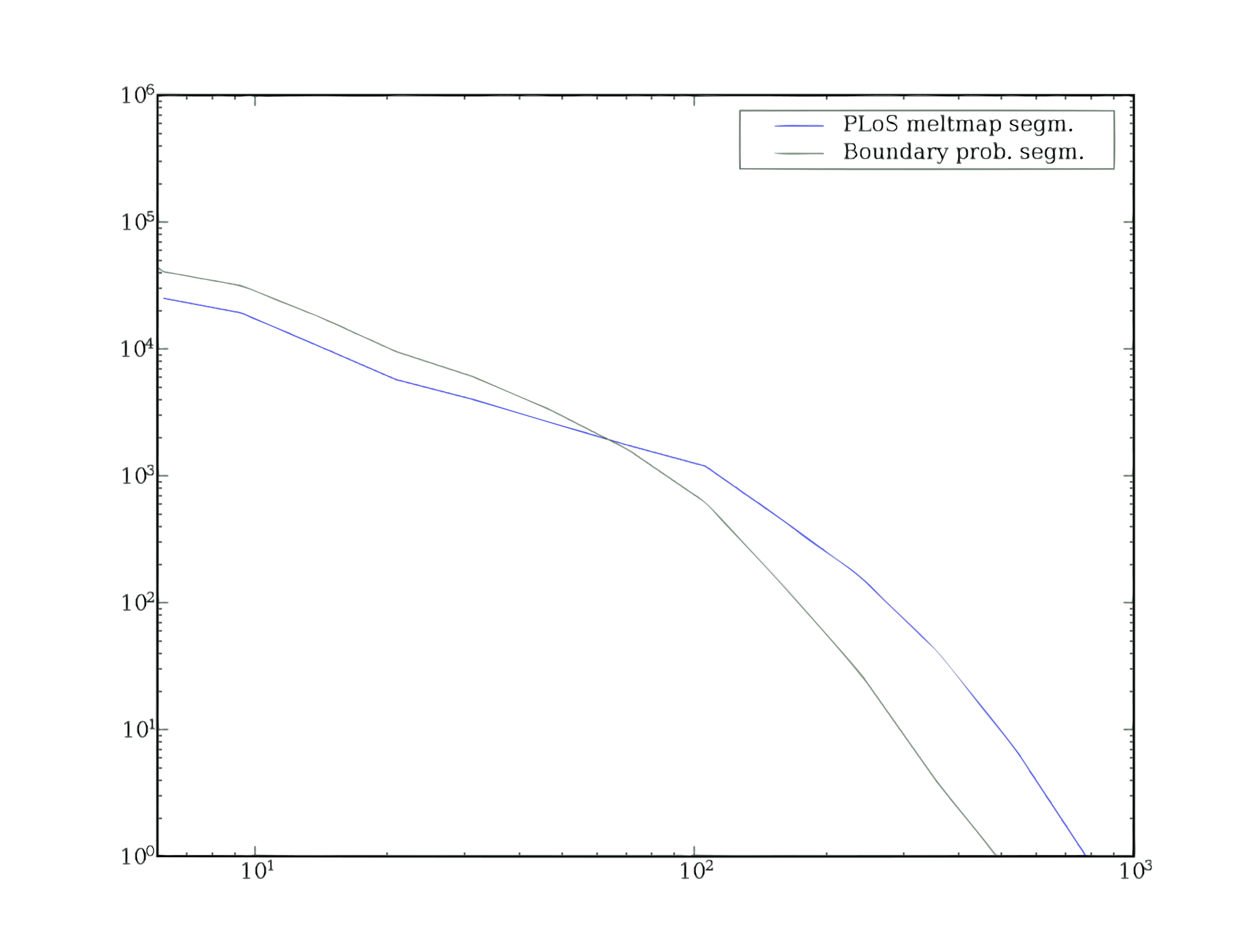} 
   \caption{Length distribution of twostate regions in human chromosome 21. This is shown using logarithmic scales, with segment length between 10 and 1000 on the x-axis, and count of segments per exact length value on the y-axis. Smoothed curves are shown for both the PLoS meltmap segmentation and the boundary probability segmentation.}
   \label{fig7}
\end{figure}

The statistical properties of the segmentations are further analyzed on human chromosome 21. \Fref{fig7}
shows a log-log plot of the length distributions of twostate/flat regions, for both the new and the previously proposed segmentation method.
Most regions are short, with relatively few regions being longer than 100 bp. 
For lengths of around 1000, there is on average less than 1 region of each exact length value. 

\begin{figure}[htbp]
   \centering
   \includegraphics[width=5in]{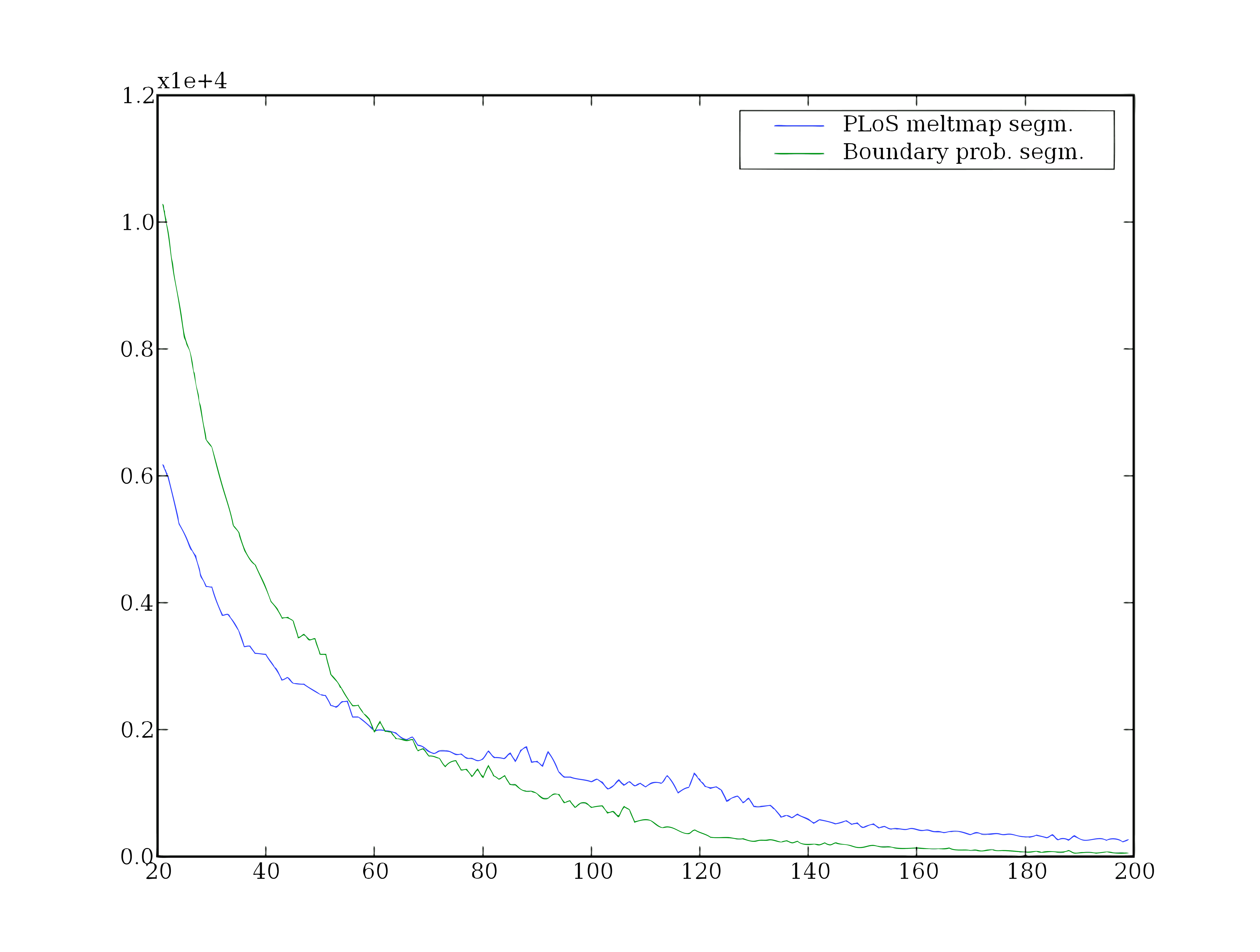} 
   \caption{Length distribution of twostate regions in human chromosome 21. This is shown using plain scales, with segment length between 20 and 200 on the x-axis, and count of segments per exact length value on the y-axis. Curves are shown for both the PLoS meltmap segmentation and the boundary probability segmentation.}
   \label{fig8}
\end{figure}

\Fref{fig8}
further shows the detailed distribution of twostate regions of length 20 to 200 using plain scales. 
From both figures it is apparent that the twostate regions of the new segmentation are generally shorter than for the previously proposed melting map segmentation.

For the full chromosome 21, there are 1.3 million twostate regions according to the new segmentation, with an average length of 18. This compares with 0.5 million flat (twostate) regions for the previous melting map segmentation, with an average length of 55.

\section{Discussion and conclusion}
In this work, we have developed a segmentation of DNA sequences
that is based on detection of local twostateness.
Azbel's groundstate segmentation produced domains that partitioned or ``tiled'' the sequence
\cite{azbel80dna-sequencing1}.
His approach ignores any excited states. In contrast, our approach takes the ensemble into account,
with the consequence that melting fork regions
appear in between the twostate regions.
Melting fork regions occupy a quite large fraction of the sequence
(\fref{fig5}).
We found that the average lengths of twostate regions are 18 bp in human chromosome 21,
while the average length of flat regions in the melting map was 55 bp
\cite{liu2007}. This is an order of magnitude smaller than previously
reported lengths of melting domains
\cite{blake87cooperative}.

These findings may have important implications for
how to utilize DNA melting in a genomic context,
in defining the extent of local DNA
singlestrandedness of a genome, and how it may
influence the informational context upon which
other molecules act.

A biproduct of the methodology is the characterization of the fluctuational noise level
in terms of the sequence dependent $\epsilon_{\text{min}}$.
From a few sequences studied so far, we have computed values in the range $10^3$--$10^4$.
Furthermore, we have found this noise level
to vary less than one order of magnitude over the melting range of temperatures,
being roughly the same at high and low temperatures.

Melting fork regions have a preferred direction: either 01 or 10.
Although a few double fork regions may actually be physically meaningful,
we suggest avoiding double fork regions by using $\epsilon\geq \epsilon_{\text{min}}$,
in order not to mistake noise for a signal.

We have aimed at generality with respect to the statistical mechanical
DNA melting model. But so far, we have only applied the methodology to the Poland-Scheraga model.
It remains to be investigated how results would depend on using other models,
and if it would be computationally viable at all. It could be a way of comparing
different DNA melting models to compute a segmentation with each.

The generality criterion imposed
some algorithmic restrictions, for example, that we could not exploit
properties of the Poland-Scheraga partition functions or the nearest neighbour
stability parameterization. It is possible that a specialized treatment
would be able to exploit such detailed knowledge to obtain much more
efficient segmentation algorithms.

The temperature scanning algorithm is simple and has algorithmic complexity $O(N)$,
but we found that small stepsizes ($\Delta T=0.1^{\circ }\text{C}$)
are needed to obtain sufficient accuracy.
Unfortunately, this means long computation times for human genomic sequences.
In this work, we have not concentrated further on algorithm optimization.

As a possible application, the boundary probability segmentation
may provide annotations of all the twostate regions
and melting fork regions in the human genome, as we did in Chromosome 21.
\emph{Regions} is a type of annotation that is well accommodated
by existing genome browsers. For such applications, it would
be useful to add an attribute to the twostate regions indicating their
(average) melting temperature.
Compared with a standard melting map, such a \emph{melting domain map}
emphasizes more the essential features of cooperativity,
while reflecting to a lesser degree the variation in GC content.
\ack{We thank Morten Johansen and Geir Ivar Jerstad.}
\appendix
\section{Calculation of boundary probabilities in the Poland-Scheraga model}
The two boundary probabilities can be computed from partition functions in
the Poland-Scheraga model as follows:
\begin{eqnarray}
p_{01}(x)=\frac{Z_{\text{X01}}(x)Z_{\text{01X}}(x)}{Z}\\
p_{10}(x)=\frac{Z_{\text{X10}}(x-1)Z_{\text{10X}}(x-1)}{Z}.
\end{eqnarray}
$Z_{\text{X01}}(x)$ and $Z_{\text{X10}}(x-1)$ are partition functions of the segment $[1,x]$,
$Z_{\text{01X}}(x)$ and $Z_{\text{10X}}(x-1)$ are partition functions of the segment $[x-1,N]$,
and $Z$ is the total partition function of the whole chain
\cite{tostesen:061922}.
All these partition functions are computed in
our previously described algorithm \cite{bip2003}.
In the language of \cite{bip2003}:
\begin{eqnarray}
p_{01}(x)=\frac{U_{01}^{\text{LR}}(x)V_{10}^{\text{RL}}(N+2-x)}{\beta Q_{\text{total}}}\\
p_{10}(x)=\frac{V_{10}^{\text{LR}}(x)U_{01}^{\text{RL}}(N+2-x)}{\beta Q_{\text{total}}}
\end{eqnarray}

Alternatively, one can use the Poland-Fixman-Freire algorithm
\cite{poland,FF77}. This algorithm is highly optimized for the purpose of
computing $p_1(x)$ profiles.
The PFF algorithm first computes all the conditional probabilities $p_{1|1}(x)$
(forward sweep) and then all the $p_1(x)$ (backward sweep).
From these two arrays, we readily obtain the boundary probabilities.
By \eref{condprobs} we derive
\begin{eqnarray}
p_{10}(x)&=p_{1}(x-1)p_{0|1}(x)\\
&=p_{1}(x-1)-p_{1}(x-1)p_{1|1}(x).\label{p10PFF}
\end{eqnarray}
By \eref{p01-p10} and \eref{p10PFF} we derive
\begin{equation}
p_{01}(x)=p_{1}(x)-p_{1}(x-1)p_{1|1}(x).
\end{equation}
\section*{References}
\bibliographystyle{unsrt}
\bibliography{unitedbps}     
\end{document}